\documentclass[
 reprint,
superscriptaddress,
 amsmath,amssymb,
 aps,
pra,
floatfix,
 lengthcheck,
 longbibliography
]{revtex4-1}
\usepackage{algpseudocode}
\usepackage[utf8]{inputenc}
\usepackage[T1]{fontenc}
\usepackage[version=4]{mhchem}
\usepackage{graphicx}
\usepackage{placeins}
\usepackage[usenames, dvipsnames]{xcolor}
\usepackage{url}
\usepackage[utf8]{inputenc}

\usepackage{verbatim}

\date{\today}

\begin{document}
\author{J.~Niskanen}
\email{johannes.niskanen@utu.fi}
\affiliation{University of Turku, Department of Physics and Astronomy, FI-20014 Turun yliopisto, Finland}
\author{A.~Vladyka}
\affiliation{University of Turku, Department of Physics and Astronomy, FI-20014 Turun yliopisto, Finland}
\author{J.~Niemi}
\affiliation{University of Turku, Department of Physics and Astronomy, FI-20014 Turun yliopisto, Finland}
\author{Ch.~J.~Sahle}
\affiliation{European Synchrotron Radiation Source,71 Avenue des Martyrs F-38000 Grenoble, France}

\title{Emulator-based Decomposition for Structural Sensitivity of Core-Level Spectra}

\begin{abstract}
We explore the sensitivity of several core-level spectroscopic methods to the underlying atomistic structure by using the water molecule as our test system. We first define a metric that measures the magnitude of spectral change as a function of the structure, which allows for identifying structural regions with high spectral sensitivity. We then apply machine-learning-emulator-based decomposition of the structural parameter space for maximal explained spectral variance, first on overall spectral profile and then on chosen integrated regions of interest therein. The presented method recovers more spectral variance than partial least squares fitting and the observed behavior is well in line with the aforementioned metric for spectral sensitivity. The analysis method is able to independently identify spectroscopically dominant degrees of freedom, and to quantify their effect and significance.
\end{abstract}

\maketitle
\section{Introduction}
Owing to orbital localization, core-level spectroscopies are sensitive to structure in the neighborhood of the excited atomic site. However, the dependence between the atomistic arrangement and the resulting spectra is not straightforward, which complicates the analysis of these spectra \cite{Ottosson2011,scirep2016,pre2017,VazdaCruz2019}. 
Satisfactory solution to this complexity calls for new methods, such as machine learning (ML) that may relieve the computational burden of repeated function evaluations \cite{hutson2020}. Here the inherent lightness of evaluation may, for example, help with problems involving predictions of statistical averages or prediction of spectra for new structures instead of their explicit simulation. Several ML approaches have recently been applied to spectroscopy \cite{fraenkel-1,fraenkel-2,fraenkel-3,gosh2019,Carbone2020,niskanen2021neural}, typically to emulate the relations between known molecular/atomic structures and corresponding spectra \cite{fraenkel-3,gosh2019}. The possibility to predict structural variations in the crystals using extended X-ray absorption fine structure has also been demonstrated \cite{fraenkel-2}. Moreover, prediction of X-ray absorption near-edge structure based on descriptors of the molecular structure has been recently shown with a high accuracy \cite{Carbone2020}.
\par
In this work, we turn to the question of how to apply an accurate ML emulator to the interpretation of core-level spectra in terms of the underlying atomistic structure. We develop a machine-learning-based dimensionality reduction of the structural parameter space based on most covered spectral variance, and apply the method to simulations for three types: X-ray photoelectron spectra (XPS), X-ray emission spectra (XES) and X-ray absorption spectra (XAS). To interpret the findings, we present a metric to measure spectral sensitivity to structural change and as the result we identify regions of higher and of lower spectroscopic structural sensitivity, consistently with the different methods.

\begin{figure*}[ht]
	\centering
	\includegraphics[width=0.88\linewidth]{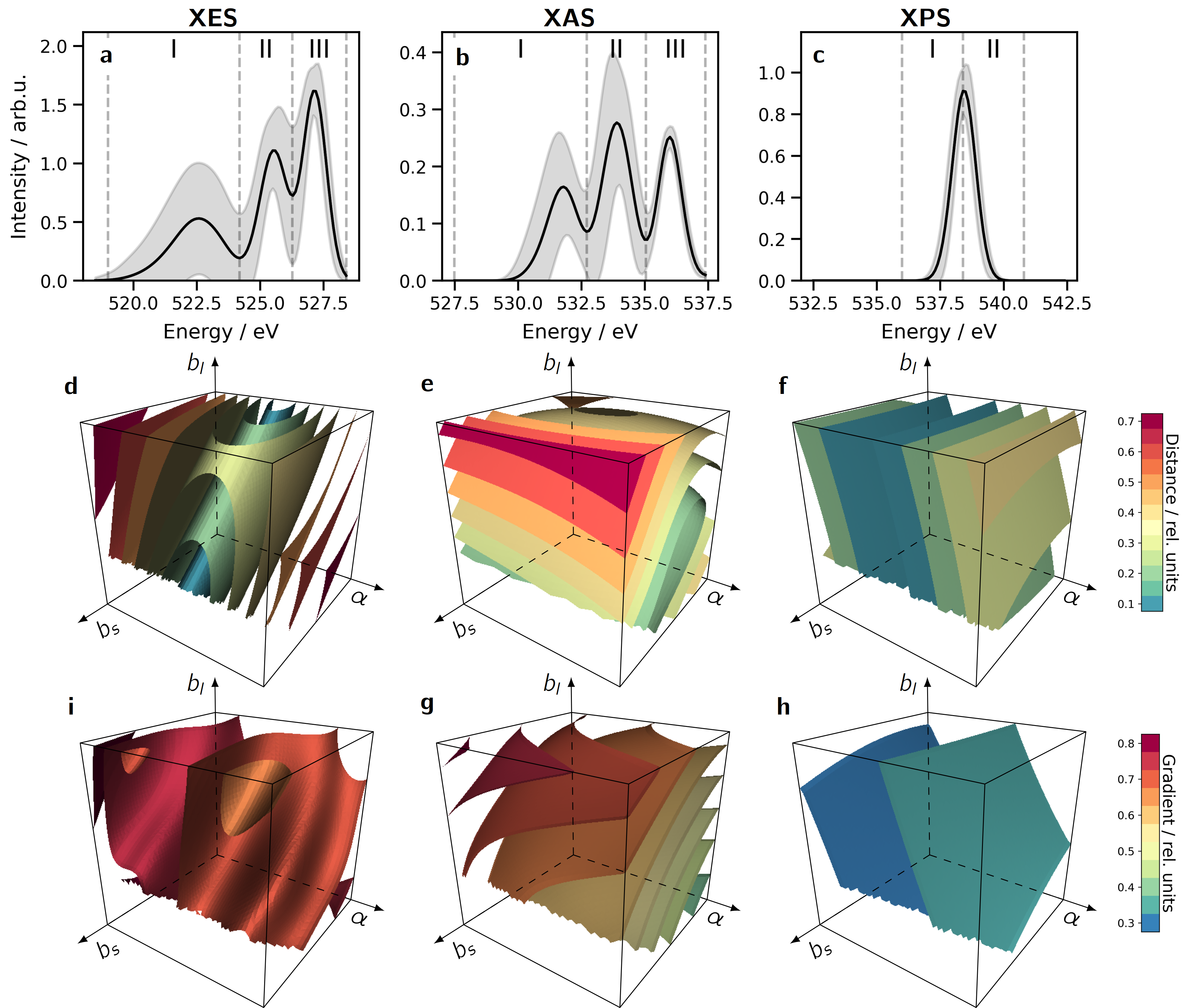}
	\caption{Spectra of the H$_2$O molecule in the training dataset; (a--c) the mean spectrum is shown in black, and the shaded area depicts $\pm 1$ standard deviation from the mean. Dashed lines indicate the regions of interests (ROIs; I, II, and III) for the coarsened spectra. Structural sensitivity of these spectra; (d--f) Cartesian distance difference $M_\mathrm{diff}$ (g--i) Jacobian norm $M_\mathrm{grad}$. Since polynomial approaches behave smoother, they were utilized also for the plots of XES. The ranges of the parameters shown are $\pm\sigma$ of the training set for parameter in question. For details, see text.}
	\label{fig:spectra}
\end{figure*}
\par
\section{Methods}
\subsection{Data and emulators}
As our data we use 10000 snapshots from {\it ab initio} molecular dynamics (AIMD) trajectories for the H$_2$O molecule with initial kinetic energy equivalent to 10000~K temperature and spectra simulated for these structures. The structural data and the related XAS spectra have been published previously \cite{niskanen2021neural}. For the calculation of XAS and XES spectra, we applied transition-potential density functional theory (TP-DFT). For evaluation of the XPS core-level binding energies, we carried out $\Delta$-DFT calculations for the hole state energy with respect to the ground state. Here we assume a high-enough photon energy to result in a constant O~1s ionization cross section regardless of the structure. All spectra were convoluted with a 1.0~eV Gaussian and are presented on a 0.1-eV-spaced grid (100 points for all cases). The calculations were carried out using the CP2K software \cite{cp2k}. The XES spectra were shifted $-6.0$~eV for easier comparison with the experiment.
\par
Our analysis relies on machine learning and the ability to predict spectra at new points in the configurational space, here defined by three degrees of freedom: H-O-H bond angle $\alpha$, the shorter and the longer O-H bond lengths $b_s$ and $b_l$, respectively. We selected the ML spectroscopic emulators in a fashion similar to that of Ref. \citenum{niskanen2021neural}. In brief, we examined polynomial models with the orders from 2 to 9, and multi-layered perceptrons (MLP) with 2--5 hidden layers and 5--500 neurons in each layer, and used mean-squared error as a metric of the training quality for a set of 8000 data points. The \texttt{scikit-learn} \cite{scikit-learn} Python package was used. 
Based on cross-validation performance scores we choose to use an MLP emulator for XES and polynomial emulators for XAS and XPS in the later stages of the analysis, carried on with a completely separate test set of 2000 samples. However, due to the wiggly behavior of the MLP isosurfaces for XES spectra, we use the smoother-behaving polynomial emulators to produce all the plots in Figure \ref{fig:spectra}.
\subsection{Spectral Sensitivity Metric}
We measure structural sensitivity as the rate of change of spectrum $\mathbf{S}(\mathbf{p})$ at structural parameter point $\mathbf{p}$. For vector-valued function $\mathbf{S}$ we define the metric
\begin{eqnarray}
\label{Eq:JacobianNorm}
    M_\mathrm{grad}(\mathbf{p})&:=&\frac{\|\mathbf{J}_\mathbf{S}(\mathbf{p})\|_2}{\|\mathbf{S}(\mathbf{p}_\mathrm{cen})\|_2}
\end{eqnarray}
where 
\begin{equation}
[\mathbf{J}_\mathbf{S}(\mathbf{p}')]_{ij}=\left.\frac{\partial S_i}{\partial p_j}\right\vert_{\mathbf{p}=\mathbf{p}'.}
\end{equation}
Each channel in the spectrum $\mathbf{S}$ is defined by the structural parameters $\mathbf{p}$. Thus each row in the Jacobian gives the gradient of the particular energy channel with respect to structure. Spectral sensitivity with respect to a given structural parameter is given by the length of the according column vector. To classify points in the configuration space, we focus on the square norm of the whole Jacobian matrix. Since we compare different spectroscopies, normalization with the spectrum at the center of the data $\mathbf{p}_\mathrm{cen}$ set is applied. 

An alternative metric is spectral deviation from that at the center of the training set
\begin{eqnarray}
\label{Eq:DifferenceNorm}
    M_\mathrm{diff}(\mathbf{p})&:=&\frac{\left\|\mathbf{S}(\mathbf{p})-\mathbf{S}(\mathbf{p}_\mathrm{cen})\right\|_2}{\left\|\mathbf{S}(\mathbf{p}_\mathrm{cen})\right\|_2}
\end{eqnarray}
Numerical calculations on a grid relied on evaluation of the ML predictor.
\subsection{Emulator-based Component Analysis}
The algorithm carries out step-wise component vector (CV) search for dimensionality reduction in the structural parameter space with the criterion to maximize the explained spectral variance together with the components of the previous steps. For a set of $N$ parameter points $\{\mathbf{p}_i\}_{i=1}^N$ this is achieved by projection on CVs optimized for the purpose. For each step $k$ ($k=1,2,...$) a unit vector $\mathbf{\hat{v}}_k$ is searched so that generalized covered variance
\begin{equation}
\rho = 1 - \mathrm{tr}(\mathbf{\Tilde{A}}^\mathrm{T}\mathbf{\Tilde{A}})/\mathrm{tr}(\mathbf{A}^\mathrm{T}\mathbf{A})
\end{equation}
is maximized. Here matrix $\mathbf{A}$ contains the true spectra of the original data points as its row vectors $\mathrm{A}_i$.  The corresponding predicted spectra for projected data points are given as row vectors of matrix
\begin{equation}
    \mathrm{A}_i^{\mathrm{(pred)}}=\mathbf{S}^\mathrm{(pred)}\left(\sum_{j=1}^k(\mathbf{\hat{v}}_j\cdot\mathbf{p}_i)\,\mathbf{\hat{v}}_j\right)
\end{equation}
where function $\mathbf{S}^\mathrm{(pred)}$ is machine-learning based emulator capable of predicting spectra for previously unseen structures and
\begin{equation}
\mathbf{\Tilde{A}}=\mathbf{A}-\mathbf{A}^{(\mathrm{pred})}.
\end{equation}
We apply the orthonormality constraint $\mathbf{\hat{v}}_k\cdot\mathbf{\hat{v}}_j=\delta_{kj}$ to the CVs and as the result of the procedure a set of orthonormal projection vectors is obtained so that they always maximize the generalized covered spectral variance $\rho$ up to the given order $k$. We applied an overall factor $\pm$1 for the CVs to point towards increasing intensity.
\par
The generalized covered variance $\rho$ accounts for the goodness score in the spectrum space, and is necessitated by the nonlinearity of spectrum prediction operation $\mathbf{S}^\mathrm{(pred)}$. When applied to data matrix from a projection in the same linear space, the definition reduces to that of covered variance used for example in principal component analysis. Due to its definition, $\rho$ may obtain negative values for notably bad predictions as the value zero corresponds to errors with the magnitude of the variance of the known data. We see no problem in alternatively using the remaining unexplained spectral variance $1-\rho$ as error metric in a minimization problem for vectors $\mathbf{\hat{v}}_k$.

\subsection{Partial least-squares fits using SVD}
We adapted an approach based on singular value decomposition (PLSSVD) \cite{booksteinPLSSVD1996} owing to its straightforward simplicity and to orthogonality of the CVs. Here the partial least-squares fit was applied to data in matrices $\mathbf{X}$ and $\mathbf{Y}$ that contain standardized structural parameters and the corresponding standardized spectra in their row vectors. A linear fit was applied between the component scores of left and right eigenvectors for each order of the decomposition. As a result, an approximation of data
\begin{equation}
    \label{eq:plssvd}
    \mathbf{Y} \approx \mathbf{X}\sum_{j=1}^{k} U^{(j)} c_j V^{(j)\mathrm{T}}
\end{equation}
was obtained. In the equation $U^{(j)}$ and $V^{(j)}$ denote the left and right eigenvectors (columnvectors) corresponding eigenvalue $\lambda_j$ ordered in descending fashion. As the data have been standardized in each of their dimensions, the covariance matrix reads directly
\begin{equation}
    \mathrm{cov}(\mathbf{X},\mathbf{Y})=\mathbf{X}^\mathrm{T}\mathbf{Y} = \mathbf{U}\,\mathrm{diag}(\lambda_1,...,\lambda_{k})\,\mathbf{V}^\mathrm{T}
\end{equation}
from which the matrices $\mathbf{U}$,$\mathbf{V}$ and $\mathrm{diag}(\lambda_1,...,\lambda_{k})$ are obtained by singular value decomposition. The procedure thus gives basis vectors on which to project the data $\mathbf{X}$ and $\mathbf{Y}$.
\par
The coefficients $c_j$ were obtained from a linear least-squares fit between between projected data points $\mathbf{X}U^{(j)}$ and $\mathbf{Y}V^{(j)}$ for each order $j=1,2,...$. The constant term in the fits was negligible and the first order coefficient is assigned $c_j$. As an example, the results of the fits for the overall spectrum case are depicted in Supplementary Information. For comparison of the PLSSVD fit results, generalized explained variance metrics were evaluated for decompositions cumulatively incremented up to order $k$ as given by Equation (\ref{eq:plssvd}). An overall factor $\pm$1 was applied for the PLSSVD structural space basis vectors to point towards increasing intensity.
\section{Results and discussion}
Although static classical nuclei model is used, the appearance of the studied spectra of the H$_2$O molecule in Figures \ref{fig:spectra} (a--c) are in agreement with the respective experiments \cite{anotherpetterssonnilssonreview2016}. The emulators trained on the sampled AIMD structures and corresponding spectra allow for easy and computationally light evaluation of the data on a mesh grid. We applied this capability to calculate the square norms of spectral deviation from that of the mean structure are depicted in Figures \ref{fig:spectra} (d--f). In addition, numerical differentiation of an emulator for the spectrum $\mathbf{S}(\mathbf{r})$ is a computationally light task on a mesh grid. Here, each partial derivative gives the rate of change for each channel in a spectrum $\mathbf{S}(\mathbf{r})$ at point $\mathbf{r}$ with respect to each structural parameter. The square norms of the Jacobian matrices $[\mathbf{J}_\mathbf{S}(\mathbf{r}')]_{ij}=\partial S_i/\partial r_j \vert_{\mathbf{r}=\mathbf{r}'}$, presented in Figures \ref{fig:spectra} (i--h) indicate strongest spectral changes in specific directions for each method. Normalization by the spectrum at the mean structure $\mathbf{r}_\mathrm{cen}$ is applied in both cases to allow for a direct comparison.
\par
The spectra show differing structural behaviour with more variation in XES and XAS than XPS, also indicated by the channel-wise one-standard deviation drawn together with the spectra. Figures \ref{fig:spectra} (e) and (g) reveal that XAS is most sensitive to the symmetric stretch. This is seen as the largest isovalue surface being located at large $b_l$ and $b_s$ values, with little variation along the bond angle $\alpha$. On the other hand the XPS spectrum changes most at high bond angles, as seen in Figures 1 (f) and (h): isosurfaces are oriented parallel to the $b_l$-$b_s$ plane. From this view XES is expected to be most sensitive to all structural parameters in the system, being least affected by the asymmetric stretch as seen in Figures \ref{fig:spectra} (d) and (i). Here the the cartesian distance difference has a low-value isosurface region intersecting the plot of Figure \ref{fig:spectra} (d), but the overall rate of change has still high isosurface values throughout the plot of Figure \ref{fig:spectra} (i).
\par
\begin{figure}[ht!]
	\includegraphics[width=\linewidth]{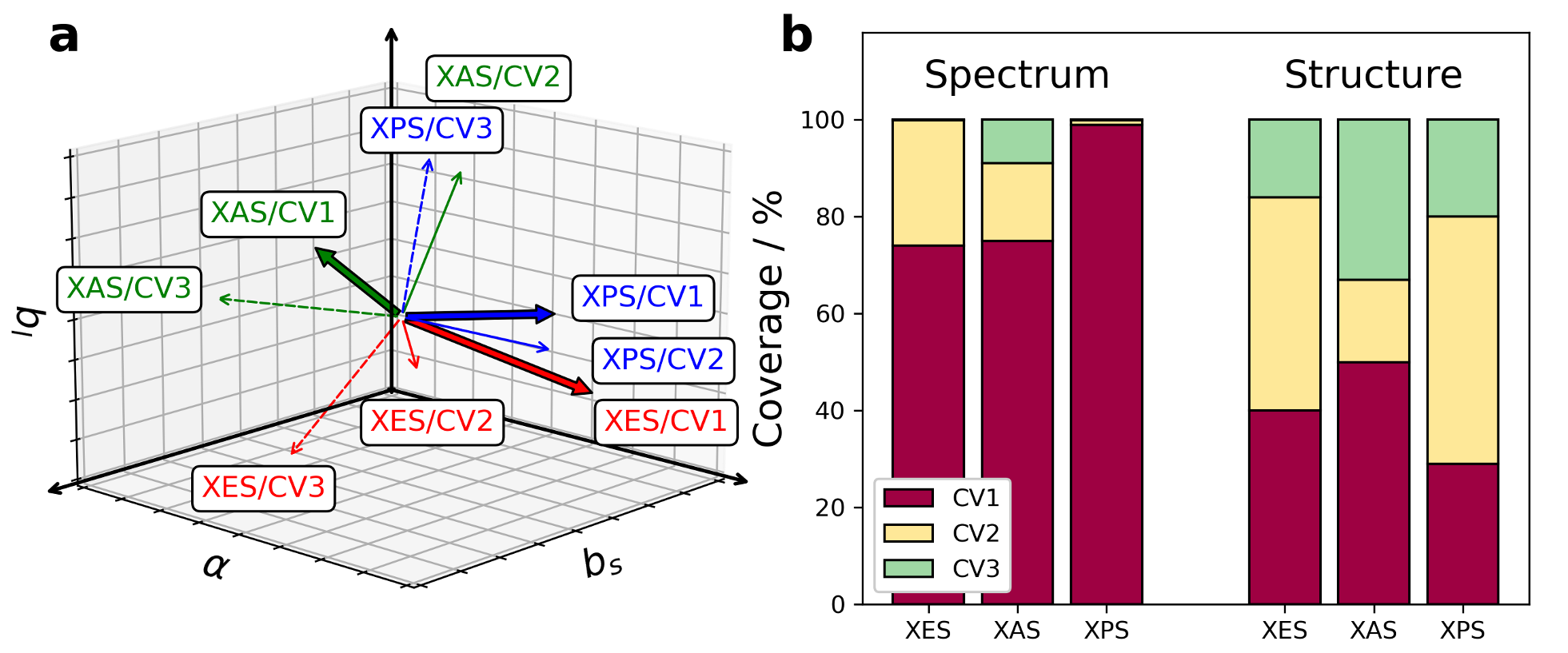}
	\caption{ECA of the full spectra. (a)~Orientation of the component vectors. Different colours indicate the type of spectroscopy, and line type depicts the components.  (b)~Ratios of explained variances for spectrum and for structure.}
	\label{fig:eca_full}
\end{figure}

\begin{figure*}[ht!]
	\centering
	\includegraphics[width=0.9\linewidth]{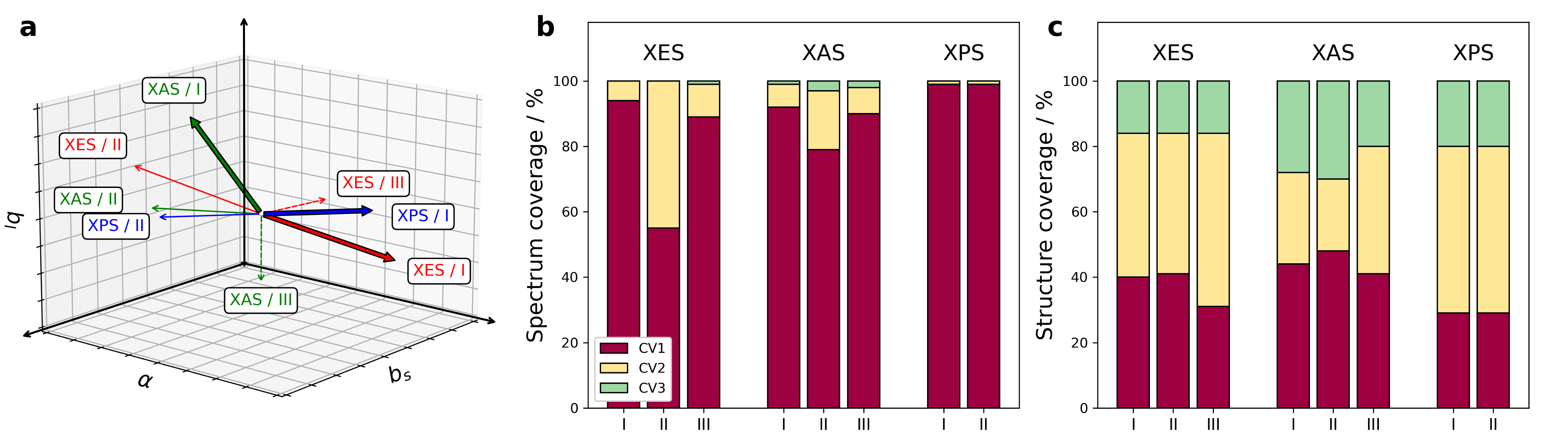}
	\caption{ROI-wise ECA of the spectra. (a)~Orientation of the first component vectors. Different colours indicate the type of spectroscopy, and line type depicts the ROI. (b)~Ratios of explained spectral variances. (c) Ratios of explained structural-parameter variances.}
	\label{fig:eca}
\end{figure*}
\par
Spectroscopic data can be seen as two correlated data sets, one for structures and one for the corresponding spectra. One way to analyse the interdependencies in such data is provided by partial least-squares (PLS) fitting \cite{geladi1988,PLSsurvey}, and a variant of this family of methods has already been applied to binding energies in XPS in aqueous solution \cite{Ottosson2011}. In PLS algorithms latent variables connecting the two data sets are searched for using only existing data points. However, we show that the relation of structure and spectra may be investigated more deeply with the help of a ML-based emulator, that is capable of making accurate and computationally light predictions of new data. Indeed, for a set of parameters defining the Hamiltonian, the spectra are defined as a function. We utilize the aforementioned capabilities of a good emulator and make a stepwise parameter-space decomposition, where the search for structural space component vectors (CV) is guided by covering of maximal variance in the spectrum space. Because the search of each CV consists of an iterative solution of an optimization problem, the lightness of evaluation of the emulator is essential. Moreover, this emulator-based component analysis (ECA) routine relies on prediction of spectra on new data {\it i.e.} projected data points in the standardized structural parameter space.
\par
When compared to the results of PLS implemented on eigenvectors from singular value decomposition of the covariance matrix (PLSSVD) \cite{booksteinPLSSVD1996}, the ECA algorithm is able to explain more spectral variance with a decomposition to a given order (Table~\ref{tab:wholespec}). Consequently, explained structural variance for ECA may be less than for the PLSSVD. We understand this by the design principle of ECA to search for directions that matter the most for spectra, with no emphasis on covered structural variance. Moreover, the nonlinearity of ECA allows for tighter match with the data than linear methods. The first CVs of the methods agree in interplay of all structural parameters, in opposing directions for angle and bond lengths for XES. Likewise, the overall shape of XAS is agreed to be dominantly affected by the bond lengths, and the XPS is virtually completely explained by the H-O-H angle. The results are also depicted in Figure~\ref{fig:eca_full} and these findings are consistent with the spectral sensitivity metrics presented in Figure~\ref{fig:spectra}.
\par
\begin{table*}[ht!]
	\small
	\centering
	\caption{Analysis of the overall shape of spectra in increasing order of decomposition: cumulative fractional explained variance in spectral ($\sigma_{spec}^2$) and structural ($\sigma_{stru}^2$) space and the corresponding component vectors in the standardized parameter space.
	}
	\label{tab:wholespec}
	\begin{tabular}{l l r r r r r r r r r r r}
		& $k$ & $\sigma_{spec}^2$ & $\sigma_{stru}^2$ & $\alpha$ & $b_l$ & $b_s$& \hspace{2em} &$\sigma_{spec}^2$ & $\sigma_{stru}^2$ & $\alpha$ & $b_l$ &$b_s$\\
		\hline
		& & ECA & & & & & & PLSSVD & & & &\\
		XES
		& 1 &  $0.74$ &  $0.41$ & [  $0.88$ & $-0.34$ & $-0.32$ ] && $0.38$ &  $0.47$ & [  $0.77$ & $-0.44$ & $-0.47$ ]\\
		& 2 &  $1.00$ &  $0.84$ & [ $-0.47$ & $-0.65$ & $-0.59$ ] && $0.51$ &  $0.85$ & [ $-0.64$ & $-0.54$ & $-0.55$ ]\\
		& 3 &  $1.00$ &  $1.00$ & [  $0.00$ & $-0.67$ &  $0.74$ ] && $0.51$ &  $1.00$ & [  $0.01$ & $-0.72$ &  $0.69$ ]\\
		XAS
		& 1 &  $0.75$ &  $0.50$ & [  $0.16$ &  $0.66$ &  $0.74$ ] &&  $0.50$ &  $0.50$ & [  $0.07$ &  $0.74$ &  $0.67$ ]\\
		& 2 &  $0.91$ &  $0.67$ & [ $-0.20$ &  $0.75$ & $-0.63$ ] &&  $0.53$ &  $0.84$ & [ $-0.98$ &  $0.17$ & $-0.09$ ]\\
		& 3 &  $1.00$ &  $1.00$ & [ $-0.97$ & $-0.05$ &  $0.25$ ] &&  $0.58$ &  $1.00$ & [  $0.18$ &  $0.65$ & $-0.74$ ]\\
		XPS
		& 1 &  $0.99$ &  $0.29$ & [  $0.96$ &  $0.26$ &  $0.03$ ] &&  $0.89$ &  $0.32$ & [ $-0.99$ & $-0.17$ & $-0.05$ ]\\
		& 2 &  $1.00$ &  $0.80$ & [  $0.14$ & $-0.42$ & $-0.90$ ] &&  $0.88$ &  $0.78$ & [  $0.17$ & $-0.93$ & $-0.33$ ]\\
		& 3 &  $1.00$ &  $1.00$ & [ $-0.23$ &  $0.87$ & $-0.44$ ] &&  $0.88$ &  $1.00$ & [  $0.01$ & $-0.34$ &  $0.94$ ]\\
		\hline
	\end{tabular}
\end{table*}
\par
Interpretation of experimental core-level spectra is complicated by unavoidable inaccuracy of the spectrum simulations. As a solution to the problem, we have previously proposed an analysis of spectral regions of interest (ROI) that are identifiable in both experiment and theory \cite{scirep2016,pre2017,jelsp2018,pnas2019,niskanen2021neural}. In such a line of thought it is argued that the risk of overanalysis is reduced, as the procedure would naturally focus on confirmedly reproduced spectral features. An alternative approach to assess uncertainties in simulated X-ray spectra have been presented by Bergmann {\it et al.} \cite{Bergmann2020}. By studying the spectral response to slight structural distortions, their method results in error bars for calculated spectra for more reliable interpretation of the experiment. 
\par
We analysed the behaviour of ROIs marked in Figures~\ref{fig:spectra} (a--c) with two approaches: simultaneous and independent for each ROI. A joint treatment of ROIs revealed that some regions dominated the component analysis at the cost of the others. This occured due to different overall variances in the ROI intensities seen in Figures \ref{fig:spectra} (a--c). For example, the optimization of the first CV became dictated by XES ROI I which resulted in highly sub-optimal description of ROI III intensity. Therefore we conclude that interpretation of ROIs is best done by individual fitting {\it i.e.} analysing each ROI separately.
\par
The results of individual analyses for each ROI are presented in Figure \ref{fig:eca} and in Table \ref{tab:roiwisefits}. When performed this way, already the first CVs explain on average
(87$\pm$14)\% of ROI intensity variance with the mean structural covered variance of
(38$\pm$7)\% as indicated by Figures \ref{fig:eca} (b--c). The first PLSSVD CVs show a weaker (68$\pm$27)\% performance for covered spectral variance 
but cover 
(42$\pm$9)\% of the structural variance. As the uncertainties above, standard deviations are given.
\par
\begin{table*}[hb]
    \small
    \centering
    \caption{Component-wise ECA analysis of the ROI intensities: cumulative fractional explained variance in spectral ($\sigma_{spec}^2$) and structural ($\sigma_{stru}^2$) space and the corresponding component vectors in the standardized parameter space. The CVs are oriented along increasing ROI intensity based on a linear fit on the predicted data for projection along the CV in question only.}
    \label{tab:roiwisefits}
    \begin{tabular}{l l r r r r r r r r r r r}
& $k$ & $\sigma_{spec}^2$ & $\sigma_{stru}^2$ & $\alpha$ & $b_l$ & $b_s$& \hspace{2em} &$\sigma_{spec}^2$ & $\sigma_{stru}^2$ & $\alpha$ & $b_l$ &$b_s$\\
\hline
    & ECA & & & & & & PLSSVD\\
XES & ROI I \\
& 1 &  $0.94$ &  $0.40$ & [  $0.90$ & $-0.31$ & $-0.32$ ] & & $0.32$ &  $0.53$ & [  $0.39$ & $-0.65$ & $-0.65$ ]\\
& 2 &  $1.00$ &  $0.84$ & [ $-0.44$ & $-0.67$ & $-0.59$ ]\\
& 3 &  $1.00$ &  $1.00$ & [ $-0.04$ &  $0.67$ & $-0.74$ ]\\
 & ROI II \\
& 1 &  $0.55$ &  $0.41$ & [ $-0.89$ &  $0.33$ &  $0.31$ ] & & $0.24$ &  $0.32$ & [ $-0.90$ & $-0.29$ & $-0.32$ ]\\
& 2 &  $1.00$ &  $0.84$ & [ $-0.46$ & $-0.62$ & $-0.64$ ]\\
& 3 &  $1.00$ &  $1.00$ & [ $-0.02$ & $-0.71$ &  $0.70$ ]\\
 & ROI III \\
& 1 &  $0.88$ &  $0.31$ & [  $0.84$ &  $0.43$ &  $0.32$ ] & & $0.69$ &  $0.36$ & [  $0.70$ &  $0.49$ &  $0.53$ ]\\
& 2 &  $0.99$ &  $0.84$ & [ $-0.53$ &  $0.62$ &  $0.57$ ]\\
& 3 &  $1.00$ &  $1.00$ & [  $0.03$ & $-0.65$ &  $0.76$ ]\\
XAS & ROI I \\
& 1 &  $0.92$ &  $0.45$ & [ $-0.42$ &  $0.88$ &  $0.25$ ] & & $0.88$ &  $0.52$ & [ $-0.38$ &  $0.76$ &  $0.53$ ]\\
& 2 &  $0.99$ &  $0.72$ & [ $-0.15$ &  $0.20$ & $-0.97$ ] \\
& 3 &  $1.00$ &  $1.00$ & [  $0.90$ &  $0.44$ & $-0.05$ ] \\
 & ROI II \\
& 1 &  $0.79$ &  $0.48$ & [ $-0.15$ &  $0.28$ &  $0.95$ ] & & $0.58$ &  $0.49$ & [ $-0.24$ &  $0.38$ &  $0.89$ ]\\
& 2 &  $0.97$ &  $0.70$ & [ $-0.14$ & $-0.95$ &  $0.26$ ] \\
& 3 &  $1.00$ &  $1.00$ & [  $0.98$ & $-0.09$ &  $0.18$ ] \\
 & ROI III \\
& 1 &  $0.90$ &  $0.42$ & [ $-0.33$ & $-0.86$ & $-0.39$ ] & & $0.80$ &  $0.51$ & [ $-0.04$ & $-0.76$ & $-0.65$ ]\\
& 2 &  $0.98$ &  $0.80$ & [  $0.92$ & $-0.20$ & $-0.33$ ] \\
& 3 &  $1.00$ &  $1.00$ & [  $0.20$ & $-0.47$ &  $0.86$ ] \\
XPS & ROI I \\
& 1 &  $0.99$ &  $0.29$ & [  $0.97$ &  $0.26$ &  $0.02$ ] & & $0.98$ &  $0.32$ & [  $0.99$ &  $0.16$ &  $0.03$ ]\\
& 2 &  $1.00$ &  $0.80$ & [  $0.13$ & $-0.38$ & $-0.92$ ] \\
& 3 &  $1.00$ &  $1.00$ & [  $0.23$ & $-0.89$ &  $0.40$ ] \\
 & ROI II \\
& 1 &  $0.99$ &  $0.29$ & [ $-0.97$ & $-0.26$ & $-0.02$ ] & & $0.98$ &  $0.32$ & [ $-0.99$ & $-0.16$ & $-0.03$ ]\\
& 2 &  $1.00$ &  $0.80$ & [ $-0.13$ &  $0.38$ &  $0.92$ ] \\
& 3 &  $1.00$ &  $1.00$ & [ $-0.23$ &  $0.89$ & $-0.40$ ]\\
\hline
    \end{tabular}
\end{table*}
\par
The CVs were oriented along the increase of corresponding ROI intensity. Whereas this is a trivial task for linear models, defining the positive direction is more complicated for ECA, because of nonlinear and possibly oscillatory behavior of intensity along the component (see Supporting Information). Our analysis reports dominant dependence on H-O-H angle of all ROIs in XES spectra: based on the first component vectors intensity transfer to ROI II is expected with inward bending. The ROIs in XAS are mostly affected by the bond lengths, and for example ROI I intensity is found to be increased with further elongation of the longer bond. Last, the sensitivity of XPS to the H-O-H bond angle only is recovered, as intensity is shifted to lower binding energies with increasing bend angles.
\par
All other things equal, a more complicated system can be expected to require a more complicated emulator architecture. This naturally will require larger training (and test) data sets that should cover the whole region of prediction \cite{niskanen2021neural}, {\it i.e.} accessible structural space. The field of machine learning provides measures how to evaluate the model and the number of required training points, by for example studying the learning curves. For the water molecule alone, a simple 3D grid evaluation would have been feasible. However, for more complicated systems the number of dimensions would prohibit such a raw approach. We see (AI)MD and Monte Carlo simulations feasible ways to generate structures as the achieved sampling cuts out large portion of the inaccessible structural space by design. These considerations are complicated by the note that the complexity of an emulator architecture depends also on how well behaving a function the spectral response is. Last, it remains a case-dependent question, how much precision loss is tolerated in the process.
\par
The idea of using decomposition is to provide interpretation for spectroscopic data learned by an emulator. The aim is to identify dominant trends in a complicated structure-spectrum relation, with inherent loss of information. In this work we used a linear transformation around a well identifiable center to identify relevant directions of spectral sensitivity. For more complicated data such as liquids, these centers may be numerous or a continuous valley of regions may appear -- possibly with varying local spectral behaviour. As one potential way to solve the problem, a manifold approach might be used. In such an approach locally linear variations would be studied together with additional parameters defining the local neighborhood, {\it e.g.} particular molecular isomer. Such parametrizations could be made by energy criteria, by abundance of points in a MD trajectory, or by principal component or a clustering analysis of the structural data. However, for spectral data that is severely wiggly or heavily scattered over the accessible structural space, it is hard to see any interpretation method to be able to draw correct universal trends from, as inverting the structure--spectrum function becomes impossible. It seems that a structural information bottleneck can be reached at least in two ways: first due to insensitivity of the probe to certain structural variation and second due to back-and-forth wiggle of the spectra in the the structural parameter space.
\par
\section{Conclusions}
Spectroscopically relevant structural variability can be captured by decomposition techniques. Utilizing ML-based emulators allows for decomposition of structural space based on explained spectral variance, an approach that outperforms partial least squares fitting both in spectral coverage and structural selectivity. The presented ECA method relies on accurate and computationally light prediction of spectra for new structures enabled by ML emulators, the development of which is currently an active field of research. Application of this analysis on ROIs in the spectrum may provide a direct interpretation for an experimentally observed and theoretically reproduced spectral change. Our results manifest X-ray spectra forming a bottleneck for structural information, some of which is not recoverable from them. Whereas high sensitivity might be beneficial for a detailed analysis of structure, sensitivity to only a few structural parameters may be utilized for identification of the related structural classes by their spectroscopic fingerprints. On the other hand, spectroscopic methods heavily sensitive on many parameters may require a statistical approach.

\section*{Acknowledgements}
Prof. E. Kukk is thanked for discussions. JN and AV acknowledge Academy of Finland for funding via project 331234.
\bibliography{references}

\clearpage
\onecolumngrid
\section*{Supporting Information}

\begin{figure}[b!]
    \centering
    \includegraphics[width=\textwidth]{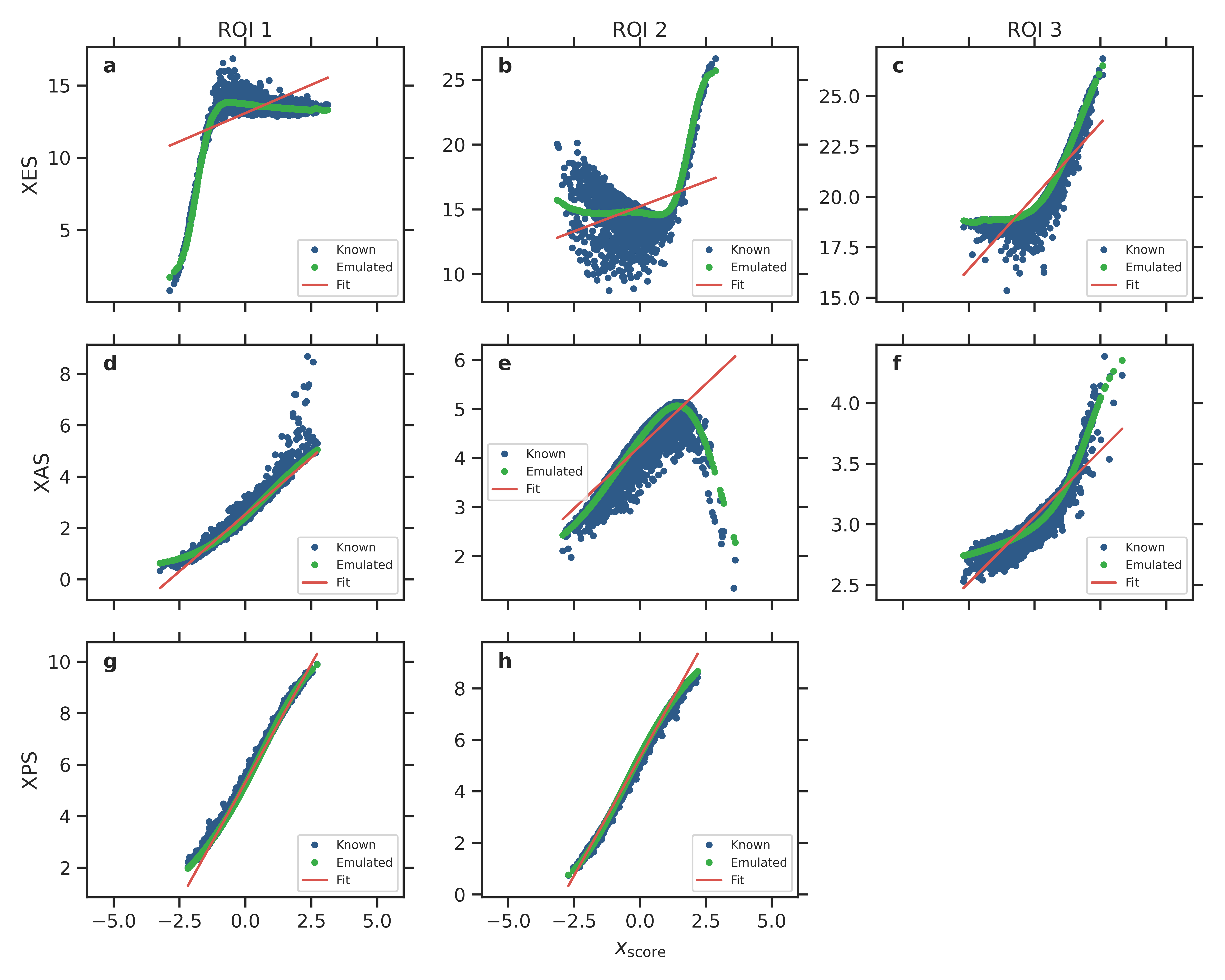}
    \caption{Linear fits to decide the orientation of the first ECA component vectors. Emulator-evaluated intensities on projected points are given together with the known intensity values for the corresponding data point.}
    \label{}
\end{figure}
\clearpage

\begin{figure}[b!]
    \centering
    \includegraphics[width=\textwidth]{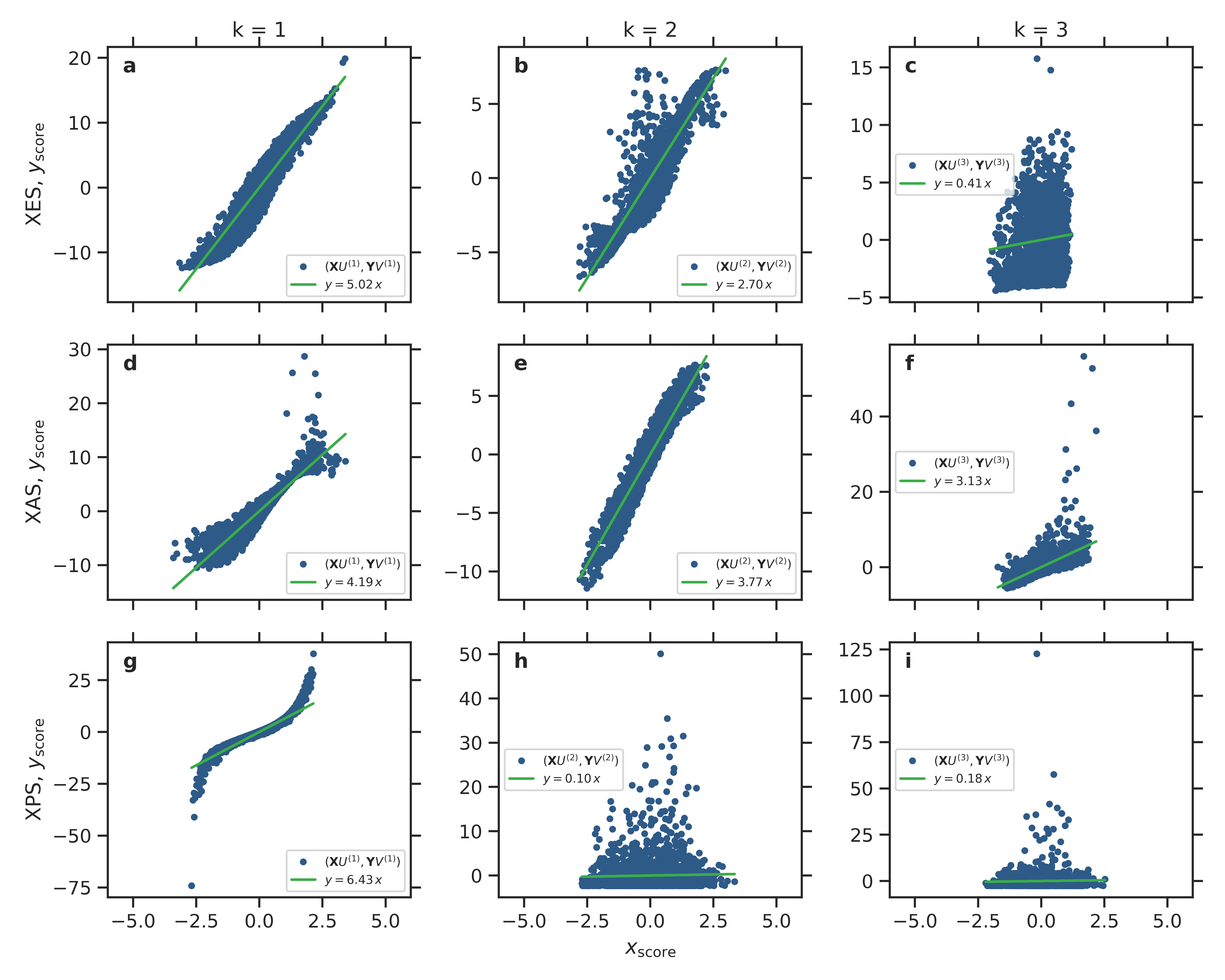}
    \caption{Fits for the coefficients $c_k$ in the PLSSVD procedure. Scores of the standardized data are given for XES (a--c), for XAS (d--f) and for XPS (g--i) together with a linear fit. The intercept term in the fit equation is negligible. }
    \label{fig:plssvd}
\end{figure}

\end{document}